\documentclass[10pt,preprint]{aastex}

\usepackage{epsfig}
\newcommand{\xperp}{ \xi_\perp }

\newcommand{\br}{\mbox{{\boldmath$r$}}} 
\newcommand{\vel}{\mbox{{\boldmath$v$}}} 
\newcommand{\bx}{\mbox{{\boldmath$x$}}} 
\newcommand{\bz}{\mbox{{\boldmath$z$}}} 
\newcommand{\bn}{\mbox{{\boldmath$n$}}} 
\newcommand{\zhat}{ \hat{\bz} }
\newcommand{\rhat}{ \hat{\br} }
\newcommand{\xhat}{ \hat{\bx} }
\newcommand{\nhat}{ \hat{\bn} }

\newcommand{\kn}{ k_n }

\newcommand{\bnabla}{ \mbox{\boldmath$\nabla$} }

\newcommand{\bs}{{\bf s}}
\newcommand{\id}{{\bf d}}
\newcommand{\bk}{{\bf k}}

\shorttitle{Scattering by a flux tube}
\begin{document}
\title{$f$-mode interactions with thin flux tubes: the scattering matrix}
\author{Hanasoge, S. M.\altaffilmark{1}, Birch, A. C.\altaffilmark{2}, Bogdan, T. J.\altaffilmark{3}, \and Gizon, L.\altaffilmark{4}}
\altaffiltext{1}{W. W. Hansen Experimental Physics Laboratory, Stanford University, Stanford, CA 94305}
\altaffiltext{2}{Colorado Research Associates Division, NorthWest Research Associates, Inc., 3380 Mitchell Lane, Boulder, CO 80301}
\altaffiltext{3}{Space Environment Center, National Oceanic and Atmospheric Administration, 325 Broadway, Boulder, CO 80305}
\altaffiltext{4}{Max-Planck-Institut f\"ur Sonnensystemforschung, Max-Planck-Str. 2, 37191 Katlenburg-Lindau, Germany}

\email{shravan@stanford.edu}
\begin{abstract}
We calculate the scattering effects associated with the interaction of a surface gravity or {\it f} mode with a thin magnetic flux tube embedded in a realistically stratified medium. We find that the dominant scattered wave is an $f$ mode with amplitude and phase of 1.17\% and around $49^\circ$ relative to the incident wave, compared to the values of 0.13\% and $40^\circ$ estimated from observations. The extent of scattering into high-order acoustic $p$ modes is too weak to be accurately characterized. We recover the result that the degree of scattering is enhanced as (a) the frequency of the incident wave increases and (b) the flux tube becomes magnetically dominated. 
\end{abstract}

\section{Introduction}
A problem of considerable interest and more recently controversy relates to the influence of magnetic fields on acoustic waves in the near-surface regions of the Sun. In these sub-photospheric magnetic regions, the ratio of magnetic to gas pressure could be very close to unity \citep[e.g.][]{mcintosh, jefferies}, leading to the contention that magnetic field effects are systematic and significant. Mode conversion \citep[e.g.][]{spruit,cally00}, is a phenomenon commonly invoked to describe magneto-hydrodynamic (MHD) wave interactions; from one acoustic mode to another, from acoustic modes to Alfv\'{e}n waves and so on, it is estimated that acoustic energy is redistributed, contributing perhaps to $p$-mode absorption observed in sunspots \citep{braun95}. 

Thin flux tubes are important building blocks of solar magnetic activity, intimately tied to studies of the solar photospheric dynamo \citep[e.g.][]{cattaneo} and flux emergence \citep[e.g.][]{cheung}. One way of deconstructing the structure and dynamics of thin flux tubes is through experimental analyses of changes in the statistics of the wave field in the vicinity of these magnetic features. 
 Applying techniques of time-distance helioseismology \citep{duvall93}, \citet{duvall06} estimated the phase and amplitude of the scattered $f$ modes from observations of thousands of isolated, small magnetic features. These parameters contain the scattering properties of an `average' flux tube, giving us insights into the nature of these flux tubes. 

However, one must apply caution when invoking causative mechanisms to explain these statistics, taking care to understand the nature of wave scattering in the presence of magnetic fields \citep[e.g.][]{spruit,bogdan95,bogdan96,cally00}. Once a sound theoretical basis is formed, observations characterizing the extent of scatter can be used to place constraints on theories of the structure and dynamics of flux tubes.

In an attempt to construct an analytical framework to decipher the nature of wave interaction with magnetic regions, \citet{bogdan96} model a magnetic flux concentration in the thin tube limit \citep[e.g.][]{spruit91} embedded in a truncated polytrope. The flux tube is assumed to be thin enough that variations across its interior are neglected; moreover, the tube must be a sub-wavelength feature and the radius small in comparison to the pressure scale height. The parameters of the polytrope are selected so as to mimic the solar sub-photospheric layers as closely as possible and is truncated a little below the photosphere because the rapidly diminishing density makes the thin tube approximation inapplicable very close to the surface. These simplifying assumptions reduce the generality of the results presented here and in fact, direct numerical solvers might appear to be a more attractive proposition. However, while numerical simulations can undoubtedly be employed to address a broader range of questions, it is also important to develop an analytical treatment of a comparable situation, as we attempt to do in this paper. Such results serve not only as benchmarks for direct computations of these phenomena but also assist us in developing a theoretical appreciation of MHD-wave interactions.

\citet{bogdan96} have demonstrated that the dominant flux tube mode is the kink mode and that the $f$ mode couples strongly with the flux tube. Moreover, the measurements of \citet{duvall06} are for $f$ modes. Consequently, the focus of this paper will be to quantify the interaction between an incoming {\it f} mode and the resulting kink mode exhibited by the flux tube.  In $\S$ \ref{mod}, we describe the model in greater detail, followed by a discussion of the method we apply to extract the scattering coefficients in $\S$ \ref{coeffs.num.sec}. The theoretical values of the phase and amplitude of the scattered $f$ mode are compared with those obtained by \citet{duvall06} (also see appendix~\ref{howtogetit}). We also demonstrate that the scattering process is predominantly restricted to $f$-$f$ while the $f$-$p$ conversions are very weak. We conclude with a summary of our calculations and a discussion of the relevance and importance of these results in $\S$ \ref{conclusions.sec}.

\section{The model\label{mod}}
The background structure in this calculation, adapted from \citet{bogdan96}, is an adiabatically stratified, truncated polytrope with index $m=1.5$, gravity ${\bf g} = - 2.775 \times 10^4~{\rm cm~s^{-2}}{\zhat}$, reference pressure $p_0 = 1.21 \times 10^5~{\rm g~cm^{-1}~s^{-2}}$, and reference density $\rho_0 = 2.78 \times 10^{-7}~{\rm g~cm^{-3}}$, such that the pressure and density variations are given by,
\begin{equation}
p(z) = p_0\left(-\frac{z}{z_0}\right)^{m+1},
\label{back.pressure}
\end{equation}
and
\begin{equation}
\rho(z) = \rho_0\left(-\frac{z}{z_0}\right)^{m}.
\label{back.density}
\end{equation}
We utilize a right-handed cylindrical co-ordinate system in our calculations, with coordinates ${\bf x} = (r,\theta,z)$ and corresponding unit vectors $(\rhat,\hat{\mbox{\boldmath $\theta$}},\zhat)$. The photospheric level of the background model is at $z=0$, with the upper boundary placed at a depth of $z_0 = 392~{\rm km}$. Following \citet{cally00}, we introduce a lower boundary at a depth of 98 Mm. The displacement potential $\Psi({\bf x},t)$ describing the oscillation modes ($t$ is time) is required to enforce zero Lagrangian pressure perturbation boundary conditions at both boundaries. This upper boundary condition is reflective in nature and therefore, possibly not very realistic. In comparison, the formalism that \citet{crouch} adopt in treating upward propagating waves at the boundary appears to be a more fitting description but is rather difficult to implement. While the choice of `good' boundary conditions is an important issue, that of incorporating the effects of the ubiquitous magnetic field that pervades the atmospheric layers is an equally, if not more important aspect \citep[e.g.][]{crouch, hanasoge07}. Moreover, the thin flux tube approximation rapidly breaks down as one proceeds higher into the atmosphere. As a first step, we have therefore chosen to (1) truncate the background model a little below the photosphere and (2) adopt simplistic boundary conditions, thereby not having to deal with the uncomfortably complex physics of the atmosphere. The incoming $f$-mode, a plane wave, which expanded in cylindrical coordinates \citep[e.g.][]{bogdan89, gizon06} has a displacement eigenfunction, $\Psi_{\rm inc}$, of the form:
\begin{equation}
\Psi_{\rm inc} = \sum_{m=-\infty}^{\infty} i^m J_m(k_0^pr) \Phi_p(\kappa_0^p;s) e^{i [m\theta -\omega t] },\label{incident}
\end{equation}
where,
\begin{equation}
\Phi_p(\kappa_n^p;s) = s^{-1/2-\mu} N_n\left[\zeta^p_n M_{\kappa^p_n,\mu}\left(\frac{s \nu^2}{\kappa^p_n}\right) + M_{\kappa^p_n,-\mu}\left(\frac{s \nu^2}{\kappa^p_n}\right) \right].\label{modes.eigfunc}
\end{equation}
In equations~(\ref{incident}) and~(\ref{modes.eigfunc}), we have introduced a slew of new symbols,
\begin{equation}
\mu = \frac{m-1}{2}, ~~~\nu^2 = \frac{m\omega^2 z_0}{g}, ~~~k^p_n = \frac{\nu^2}{2\kappa^p_n z_0},\label{eig.values}
\end{equation}
$\omega$ the angular frequency of oscillation, $s = -z/z_0$, $J_m(w)$, the Bessel function of order $m$ and argument $w$ and $M_{\kappa, \mu}(w)$, the Whittaker function \citep[e.g.][]{whittaker} with indices $\kappa, \mu$ and argument $w$. The eigenvalue $\kappa^p_n > 0$ and constant $\zeta^p_n$ characterizing the mode are obtained through the procedure described in appendix~\ref{append.eigvalues}.
The $n=0$ mode corresponds to the surface gravity or $f$ mode, while $n > 0$ represents the acoustic $p_n$ mode. The term $N_n$ is the normalization constant for the mode, defined as
\begin{equation}
N_n = \left[\int_1^{\infty}\left[\zeta^p_n M_{\kappa^p_n,\mu}\left(\frac{\nu^2 s}{\kappa^p_n}\right) +M_{\kappa^p_n,-\mu}\left(\frac{\nu^2 s}{\kappa^p_n}\right) \right]^2 ds \right]^{-1/2}.
\end{equation}

\subsection{Flux tube}
Applying the approximations listed in $\S$2 of \citet{bogdan96}, a thin flux tube carrying a magnetic flux of $\Phi_f = 3.88 \times 10^{17} {\rm Mx}$, with plasma-$\beta =1$ everywhere inside the tube is placed in the polytrope. The thin flux tube approximation,
\begin{equation}
b(s) \approx \sqrt\frac{8\pi p(s)}{1+\beta}, ~~~~\pi R^2(s) \approx \frac{\Phi_f}{b(s)},
\end{equation}
where $b(s)$ and $R(s)$ are the magnetic field and the radius of the tube at depth $s$, is shown to be accurate to better than a percent in the truncated polytrope situated below $z=-z_0$ or $s=1$ \citep{bogdan96}. Note that the magnetic flux associated with the tube is held constant - different values of $\beta$ therefore result in different $b(s)$ and $R(s)$.
\subsection{Oscillations of the tube: the kink mode}
Horizontal motions of the flux tube created by the impinging modes (whose displacement potential is given by $\Psi({\bf x},t)$) in the direction of the wave vector are described by $\xi(s,t)$ \citep[e.g.][]{bogdan96}, a solution to the differential equation
\begin{equation}
\left[z_0\frac{\partial^2}{\partial t^2} - \frac{2gs}{(1+2\beta)(m+1)}\frac{\partial^2}{\partial s^2}- \frac{g}{1+2\beta}\frac{\partial}{\partial s}\right]\xi = \frac{2(1+\beta)}{1+2\beta} z_0\frac{\partial^3\Psi}{\partial x\partial t^2},\label{xiperp}
\end{equation}
where $x = r\cos\theta$. Following \citet{bogdan96}, we define $\xperp = -i\tilde{\xi}(s)$, where $\xi(s,t) = \tilde{\xi}(s)e^{-i\omega t}$, and $\tilde{\xi}(s)$ is the purely spatial component of the tube displacement. The function $\xperp$ contains all of the scattering information that is needed to understand the interaction of the wave field with the flux tube. 

\subsection{Jacket modes and a lower boundary\label{jacket.exposition}}
\citet{bogdan95} showed that scattered waves created as a consequence of magnetic interactions are a mixture of modal and evanescent components. The scattering process results in not only a redistribution (and loss) of modal energies but also in the production of a continuous spectrum of evanescent `modes' called {\it jacket modes}. The existence of evanescent waves in the near field of an arbitrary scatterer (not just magnetic) that possesses subwavelength features has been well documented in the area of acoustics \citep[e.g.][]{maynard} and optics \citep[e.g.][]{boz}. Mathematically, these evanescent modes appear to compensate for the inability of the set of propagating-mode eigenfunctions to represent a spatially intricate scatterer. These non-propagating evanescent waves decay exponentially rapidly with distance from the scatterer. 

The formalism of \citet{bogdan95} requires an uncountable infinity of jacket modes to complete the basis, a consequence of the lower boundary being placed at $s=\infty$. Jacket modes in this instance are mathematically described using Whittaker functions, which are relatively difficult and expensive to compute accurately. Moreover, the jacket mode equations listed in \citet{bogdan95} contain integrals over running indices that pose significant numerical hurdles because of the poor convergence properties of the integral. In order to circumvent any calculations involving these continuous jacket modes, we employ a lower boundary placed at a depth $s = D = 250$ ($z = 98$ Mm) to reduce this uncountably infinite set of evanescent modes to a more tractable discrete countably infinite counterpart in the manner described in \citet{cally00}.

Only modes whose inner turning points are in the vicinity of the lower boundary are affected by its presence and therefore, placing it at a depth of 98 Mm means that the $f$ and first few $p$ modes remain unaware of its presence (because of the decay of the eigenfunction). Moreover, the scattering amplitudes decrease sharply with increasing radial order. Consequently, the high order $p$ modes which interact with the lower boundary are largely unimportant in any case because of their weak contribution to the scattering process studied here. With the introduction of the boundary, the problem becomes numerically well defined as well.

 In order to obtain the $p$- and jacket mode eigenfunctions, we apply a zero Lagrangian pressure perturbation lower boundary condition for the sake of simplicity \citep[see the appendix in][]{cally00}. We also assume that the tube oscillations ($\xperp$ in Eq.~[\ref{xiperp}]) are oblivious to the lower boundary. This condition is necessary because energy loss in the form of propagating Alfv\'{e}n waves along the tube can only occur when the lower boundary is transparent (not the case with zero Lagrangian pressure perturbation).

To summarize, we use the formalism and model of \citet{bogdan96} but unable to carry out calculations of the necessary jacket modes described in \citet{bogdan95}, we replace this uncommonly difficult continuous set of modes by its more tractable discrete cousin \citep{cally00}. Next, we describe the tube radius boundary condition that allows us to begin the task of estimating the scattering coefficients.

\subsection{The scattered wave and tube boundary}
The scattered wave is given by \citep[e.g.][]{bogdan95}:
\begin{equation}
\Psi_{\rm sc} = -\sum_{n=0}^{\infty}\sum_{m=-\infty}^{\infty} i^m \left[\alpha^p_{mn} H^{(1)}_m(\kn^p r) \Phi_p(\kappa^p_n;s) + \beta^J_{mn} K_m(\kn^J r) \Phi_J(\kappa^J_n;s) \right] e^{i [m\theta -\omega t] },\label{scatter.1}
\end{equation}
where $\alpha^p_{mn}$ are the $p$-mode scattering coefficients, $\beta^J_{mn}$ are the jacket mode coefficients and $K_m(w)$ is the $K$-Bessel function of order $m$ and argument $w$. The resonant wavenumber of the $p_n$ mode is denoted by $k^p_n$ (including the $n=0$ {\it f} mode), while wavenumbers corresponding to jacket modes are labeled $k^J_n$. The eigenvalues $\kappa^J_n > 0$ and $\kappa^p_n$ are related to $k^J_n$ and $k^p_n$ respectively, according to equation~(\ref{eig.values}) (replace $p$ by $J$ in Eq.~\ref{eig.values}). In equation~\ref{incident}, The un-normalized jacket mode eigenfunction, $\Phi_J(\kappa^J_n;s)$, is given by 
\begin{equation}
\Phi_J(\kappa^J_n;s) = s^{-1/2 - \mu} \left[\eta^J_n M_{-i\kappa^J_n,\mu}\left(\frac{i\nu^2}{\kappa^J_n}s\right) + M_{-i\kappa^J_n,-\mu}\left(\frac{i\nu^2}{\kappa^J_n}s\right) \right],\label{jacket.modes}
\end{equation}
where $\eta^J_n$, a parameter and $\kappa^J_n$, the jacket mode eigenvalue are determined by the boundary conditions (see appendix A for details) and $M_{\kappa,\mu}(w)$ is the Whittaker function \citep{whittaker}. The normal to the tube boundary at a given depth $s$ is given by
\begin{equation}
\nhat = \frac{{\rhat} - \frac{1}{z_0}\frac{dR}{ds}{\zhat}}{\left[1 + \left(\frac{1}{z_0}\frac{dR}{ds}\right)^2\right]^{1/2}}.\label{eq.nhat}
\end{equation}
A consequence of the thin flux tube approximation is that $|\frac{dR}{ds}| << 1$; however we retain this term throughout in order to verify that the scattering coefficients are at best weakly dependent on it (see $\S$\ref{coeffs.num.sec}). The boundary condition is then obtained by matching radial velocities across the tube boundary, $r = R(s)$,
\begin{equation}
\nhat\cdot\bnabla\left[\Psi_{\rm inc} + \Psi_{\rm sc}\right]_{r=R(s)} = \xhat\cdot\rhat \xperp(s) e^{-i\omega t},\label{bc.1}
\end{equation}
where $\xhat$ is the unit vector along the x-axis. Simplifying equation~(\ref{bc.1}), we obtain
\begin{equation}
\left[\frac{\partial}{\partial r}(\Psi_{\rm inc} + \Psi_{\rm sc}) -\frac{1}{z_0^2}\frac{dR}{ds}\frac{\partial}{\partial s}(\Psi_{\rm inc} + \Psi_{\rm sc})\right]_{r=R(s)}  = e^{-i\omega t} \xperp(s) \cos{\theta}\left[1 + \left(\frac{1}{z_0}\frac{dR}{ds}\right)^2\right]^{1/2}.\label{bc.finally}
\end{equation}
Using a least squares approach, equation~(\ref{bc.finally}) is then solved to obtain an estimate for the scattering coefficients $\alpha^p_{mn}$.

\section{Solution procedure}\label{coeffs.num.sec}
By only retaining the $|m| =1$ coefficients and canceling the $e^{-i\omega t}$ term in equation~(\ref{bc.finally}), time and angular dependencies may be eliminated. Because we are only dealing with a single $m$, the scattering coefficients introduced in equation~(\ref{scatter.1}) are rewritten as $\alpha_n^p, \beta_n^J$. We define the following functions 
\begin{equation}
f(s) = \frac{i\xperp(s)}{2}\sqrt{1 + \left(\frac{1}{z_0}\frac{dR}{ds}\right)^2} + \left(\frac{\partial}{\partial r} - \frac{1}{z_0^2}\frac{dR}{ds}\frac{\partial}{\partial s}\right)\left[  J_1(k^p_0 r)\Phi_p(\kappa^p_0;s)\right]|_{r = R(s)},\label{eq14}
\end{equation}
\begin{equation}
g^p_n(s) = \left(\frac{\partial}{\partial r} - \frac{1}{z_0^2}\frac{dR}{ds}\frac{\partial}{\partial s}\right)\left[ H^{(1)}_1(k^p_n r) \Phi_p(\kappa^p_n;s)\right]|_{r = R(s)},
\end{equation}
\begin{equation}
g^J_n(s) = \left(\frac{\partial}{\partial r} - \frac{1}{z_0^2}\frac{dR}{ds}\frac{\partial}{\partial s}\right)\left[ K_1(k^J_n r) \Phi_J(\kappa^J_n;s)\right]|_{r = R(s)},\label{eq16}
\end{equation}
and arrive at the least-squares problem:
\begin{equation}
A\pmatrix{[\alpha]\cr[\beta]} = \pmatrix{ f(s_1)\cr...\cr f(s_M)}\label{least.squares}.
\end{equation}
In equation~(\ref{least.squares}), $s_1$ and $s_M$ ($M$ is the number of grid-points in depth) are the start and end of the finite vertical domain, $[\alpha], [\beta]$ are column vectors of the scattering coefficients, $\alpha_n^p$ and $\beta_n^J$ and the $M \times (N_1+N_2+1)$ matrix $A$ is given by: 
\begin{equation}
A = \pmatrix{g^p_0(s_1)&...&g^p_{N_1}(s_1)&g^J_{1}(s_1)&...&g^J_{N_2}(s_1)\cr...&...&...&...&...&...\cr g^p_0(s_M)&...&g^p_{N_1}(s_M)&g^J_{1}(s_M)&...&g^J_{N_2}(s_M)}, 
\end{equation}
where $N_1+1, N_2$ are the number of $p$- and jacket modes included in this calculation, respectively. The grid spacing in the $s$ space was set at $\Delta s = 0.249$ or correspondingly, $\Delta z = 97$ km. For physical parameters, we utilize the model described in \citet{bogdan96}, wherein $m=1.5$, $\Phi = 3.88\times 10^{17} ~{\rm Mx}$, $R(s=1) = 100 ~{\rm km}$, $z_0 = 392 ~{\rm km}$, and incident modes of frequency, $\omega =2\pi\nu, \nu=2, 3, 4, 5$ mHz,  and $\beta =0.1, 1, 10$. The lower boundary of the box is set at $D = 250$ or 98 Mm. The thin flux tube approximation may be invoked to ignore the dependence of various terms on derivatives with respect to $s$ in this calculation. Although not shown here, we verified the validity of this approximation by demonstrating the invariance of the scattering coefficients to the presence of the derivative terms. That the derivative terms contribute very little to the solution is an important test of the thin flux tube approximation. Also, we considered a near-field approximation of the terms $J_1(k_0^p r)$, $H_1^{(1)}(k_n^p r)$, and $K_1(k_n^J r)$ in equations~(\ref{eq14})-(\ref{eq16}). Applying this approximation to $J$ and $H$ does not significantly change our solutions.  However, the near-field approximation to the Bessel $K$ leads to substantially different solutions for the scattering coefficients.

 The coefficients for the first 4 modes were seen to be stable to changes in the depth of the lower boundary, as discussed in $\S$\ref{jacket.exposition}. To test the accuracy of the $p$-mode scattering coefficients, we attempted to match the energies of the ingoing with the outgoing and Alfv\'{e}n waves. However, because their amplitudes are so low, we find that small changes in the phase of the scattered $f$ mode result in significant changes in the amplitudes of the high-order scattered $p$ modes. This implies a lack in the robustness of the result; hence we only show the values of the $f$- and $p_1$-mode scattering coefficients.

\begin{figure}
\begin{center}
\epsfig{file=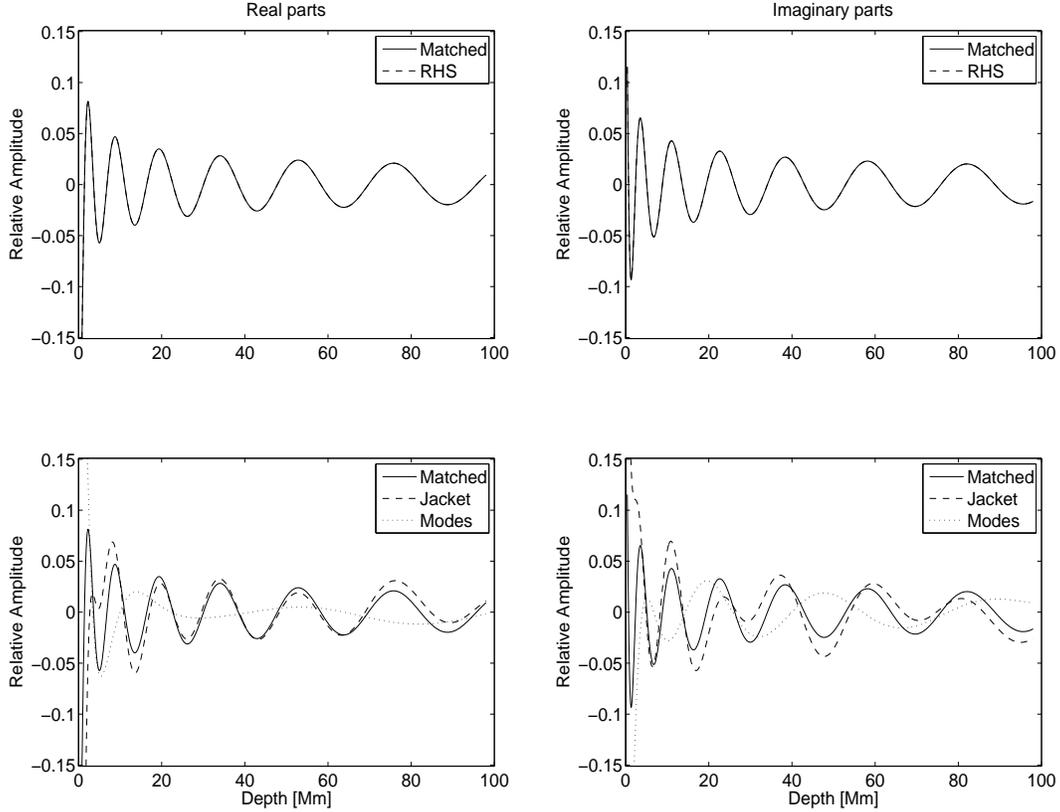,height=12cm,width=\linewidth}
\vspace{-0.5cm}
\caption{The upper panels show comparisons of real and imaginary parts of the fit (labeled `matched' in the figure) to the corresponding real and imaginary parts of the right hand side ($=f(s)$, Eq. [\ref{eq14}]) for $\beta = 1$ and $\nu = 3$ mHz. The curves are indistinguishable. Contributions to the real and imaginary parts of the right hand side (RHS) respectively from the jacket and resonant modes are shown in the lower panels. The lower boundary was placed at a depth of 98 Mm; the surface gravity mode with 8 normal and 675 jacket modes were used in this matching. The scattered amplitude is normalized with respect to the incident $f$-mode amplitude.}\label{compare.real.imag}
\end{center}
\end{figure}

\begin{figure}
\begin{center}
\epsfig{file=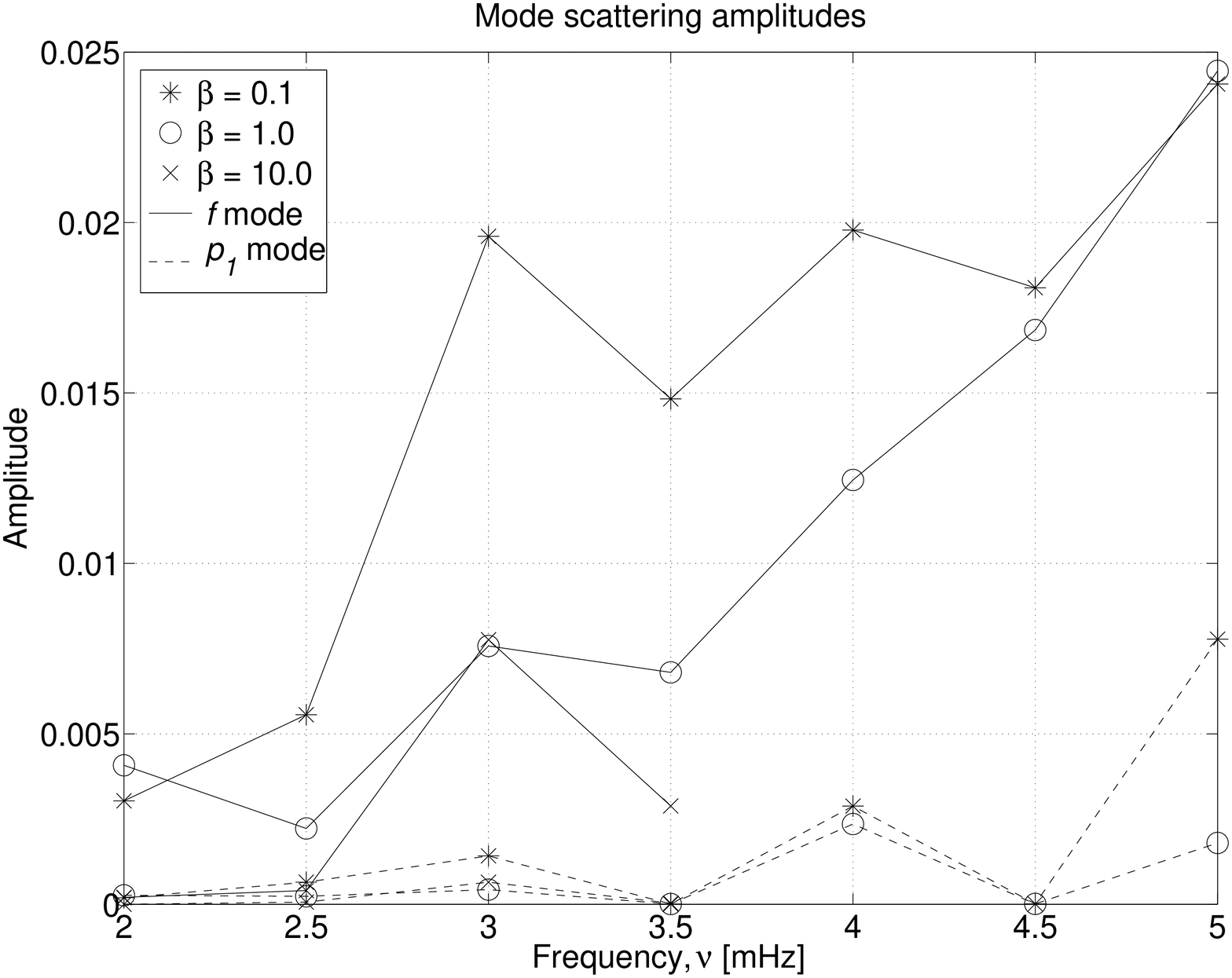,height=9cm}
\caption{The amplitudes of the scattered $f$ (solid line) and $p_1$ (dashed line) modes at various values of plasma-$\beta$ and frequency, $\nu$. The $p_1$ scattering coefficients are seen to be substantially smaller than the corresponding ones for the $f$ mode.}\label{amp_phases}
\end{center}
\end{figure}

Large numbers of jacket modes (463, 675, 820, 857 for $\nu =2, 3, 4, 5$ mHz respectively) are required to fit the right hand side with the accuracy of Figure~\ref{compare.real.imag}. Least-squares fitting was performed using the backslash command in MATLAB. The various complementary and standard Whittaker functions were computed using CERNLIB, a freely available suite of mathematical functions. The absolute values of a sample of the scattering coefficients, $|\alpha^p_n|$ and corresponding phases, $\arg(\alpha^p_n)$ for plasma-$\beta=1$ and $\nu = 3$ mHz are listed in the second and third columns of table~\ref{scatter.coeffs}. We do not display the jacket mode coefficients here.
\begin{deluxetable}{ccc}
\tablecolumns{3}
\tablewidth{0pc}
\tablecaption{Scattering coefficients\label{scatter.coeffs} for $\beta=1$, $\nu=3$ mHz for a flux tube with radius 100 km at $s=1$.}
\tablehead{\colhead{Mode} & \colhead{Amplitude} & \colhead{Phase}}
\startdata
$f$& 0.0117 &49.6$^\circ$\\
$p_1$& 0.0007 &76.7$^\circ$\\
\enddata
\end{deluxetable}
Our confidence in these values is strengthened by the consistency in the values of the scattering coefficients over a large number of numerical experiments. We show in the upper panels of Figure~\ref{compare.real.imag} that the combination of resonant and jacket modes captures the right hand side very well. In the lower panels, the contributions of the jacket and resonant modes are separated to illustrate that jacket modes are a non-trivial component of the scattering process. Shown in Figure~\ref{amp_phases} is the agreement with intuitive expectation that the higher the radial order of the $p$ mode, the lower the contribution. This is in line with the idea of a diagonally dominant scattering matrix \citep[for scattering matrix, see e.g.][]{braun95}; i.e., a scattering matrix  with weak off-diagonal terms that decay very rapidly. 

\begin{figure}
\begin{center}
\epsfig{file=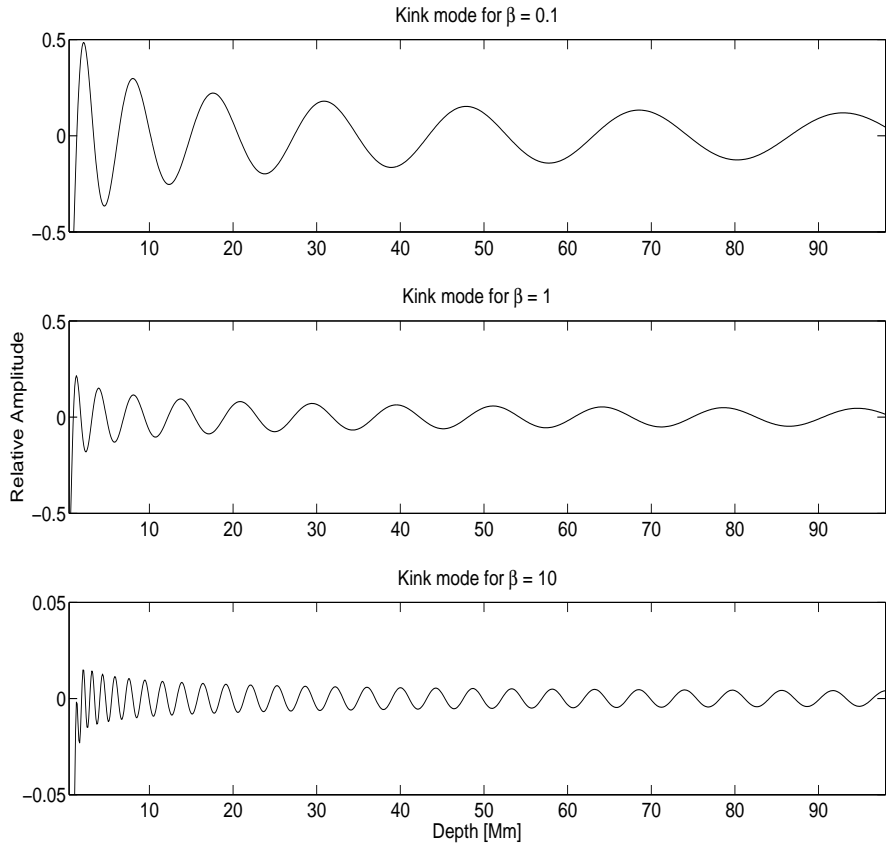,height=12.5cm,width = 14cm}
\vspace{-0.5cm}
\caption{Kink modes at $\nu = 5$ mHz for different values of $\beta$. The magnetically dominated case with $\beta = 0.1$ is seen to be very stiff, as opposed to the $\beta = 10$ case. The amplitude of the mode is normalized by the amplitude of the incoming $f$ mode.}\label{kink}
\end{center}
\end{figure}
\begin{figure}
\begin{center}
\epsfig{file=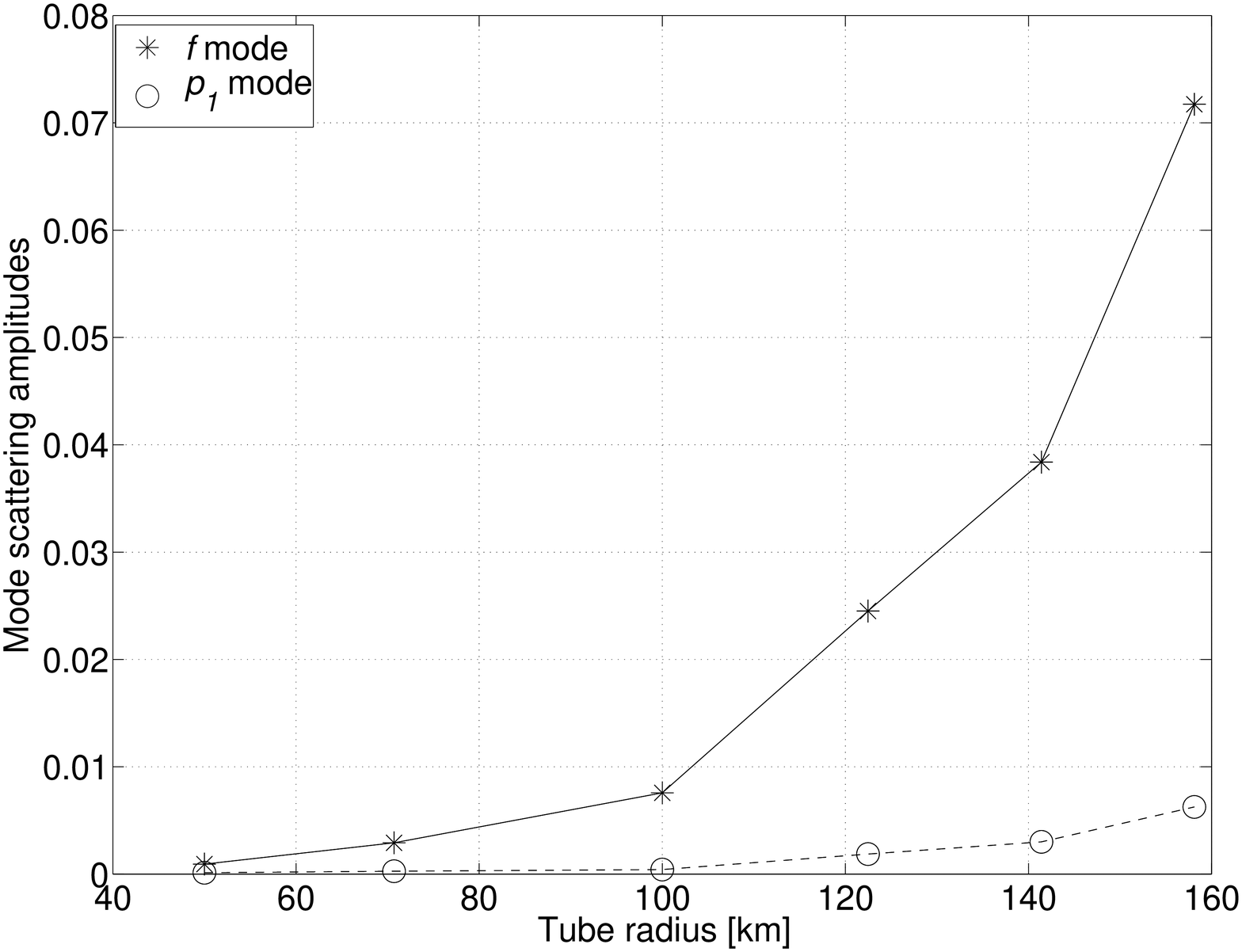,height=9cm}
\caption{Scattering coefficients for $\nu = 3$ mHz, $\beta =1$ as a function of tube radius. The flux, $\Phi \propto R^2$, where $R$ is the tube radius. A clear trend with increasing tube radius (or flux) is seen. The thin flux tube approximation requires the tube radius to be smaller than a scale height (= 260 km); therefore, we cannot study the scattering at large values of $R$.}\label{amp_radius}
\end{center}
\end{figure}

There is a definite dependence of the magnitude of the scattering coefficients on the plasma-$\beta$ and the frequency of the incident wave \citep[e.g.][]{bogdan96}. It emerges from our calculations that the wave - flux tube coupling becomes stronger as the frequency increases (see Figure~\ref{amp_phases}). Moreover, as the flux tube becomes magnetically dominated ($\beta = 0.1$), the scattering is enhanced, as emphasized by the magnitudes of the scattering coefficients. Correspondingly, the flux tube becomes relatively stiff and the tube kink mode is more resistant to the buffeting forces of the interacting modes (upper panel of Figure~\ref{kink}). The reverse effect is observed for $\beta = 10$, where the tube is hydrodynamically dominated (lowest panel of Figure~\ref{kink}). Also, the scattering coefficients increase non-linearly with the amount of flux (and therefore tube radius) at fixed $\beta$ and $\nu$. Since the thin flux tube approximation requires that the tube radius at the surface be smaller than a scale height, we can only study a restricted set of radii (see Figure~\ref{amp_radius}).


\section{Conclusions}\label{conclusions.sec}
We have presented a model of wave scattering off a magnetic flux concentration. Earlier efforts to model magneto-acoustic interactions \citep[e.g.][]{gizon06} have been under less realistic conditions because of the related mathematical difficulties. However, it is noted that this model does not capture the complexities of the solar case where phenomena such as radiative heat transfer, azimuthal asymmetries in the tube, the lack of applicability of the thin tube approximation in near-surface regions, local changes in source properties, downflows, etc. probably play an important role in affecting the wave field. Even under relatively idealized conditions as in this model, the manipulations required to recover the scattering properties are non-trivial. Perhaps this indicates that the formalism is sub optimal when keeping in mind the relatively significant effort required to compute in this basis set of Whittaker functions while coping with the restricted generality of the solutions. With the development of methods of computational forward modeling to address issues of wave interactions with magnetic fields \citep[e.g][]{cameron07,hanasoge07} comes the ease of studying these problems in greater generality. The estimates for the degree of scattering obtained here can be used to test the validity of numerical forward calculations.

It is shown that the contribution to the scattering process by the so-called {\it jacket} or evanescent modes is entirely non trivial, and therefore inevitable in studies of scatterers that contain subwavelength variations in their spatial structure. It has been realized by \citet{maynard} that these near-field evanescent modes may be analyzed to image with subwavelength resolution. Thus, future prospects of studying these flux tubes in greater detail are exciting, especially in the context of upcoming satelite missions. 

Lastly, to say something about the nature of the average flux tube in the measurements of \citet{duvall06} based on this analysis, we first assume that the interactions are well captured by the single scattering approximation. This implies that the scattering magnitude will scale linearly with the number of model flux tubes while the phase of the scattered wave is unchanged from the value listed in Table~\ref{scatter.coeffs}. While our scattering model over estimates the amplitude of scatter by a factor of 8.8, the phase of the  $f$-mode scattering coefficient agrees well with the observations. Several possibilities arise: either the $\beta \gg 1$ for flux tubes in the Sun, or the upper boundary conditions of this model are inaccurate, or there are important pieces of physics, like non-linearities, radiative heat transfer etc., not accounted for here. More conclusive results await further investigations.

\acknowledgements
S.M.H. was funded by NASA grants MDI NNG05GH14G and HMI NAS5-02139. Computations were performed on the Stanford Solar group machines.

\appendix
\section{Eigenvalues\label{append.eigvalues}}
\subsection{Jacket mode eigenvalues\label{jack.sec}}
 The functional form of jacket modes used in our calculations is given by equation~(\ref{jacket.modes}). These functions are forced to satisfy the boundary conditions,
\begin{equation}
\frac{\partial\Phi_J}{\partial s} + \frac{\nu^2}{m}\Phi_J = 0
\end{equation}
at $s=1$ and $s=D$. Defining 
\begin{equation}
N_J(\kappa^J_n,\mu,s) = \frac{1/2 + \mu}{s} M_{-i\kappa^J_n,-\mu}\left(\frac{i\nu^2}{\kappa^J_n}s\right) - \frac{i\nu^2}{\kappa^J_n}M'_{-i\kappa^J_n,-\mu}\left(\frac{i\nu^2}{\kappa^J_n}s\right) -\frac{\nu^2}{m} M_{-i\kappa^J_n,-\mu}\left(\frac{i\nu^2}{\kappa^J_n}s\right),
\end{equation}
and
\begin{equation}
D_J(\kappa^J_n,\mu,s) = -\frac{1/2 + \mu}{s} M_{-i\kappa^J_n,\mu}\left(\frac{i\nu^2}{\kappa^J_n}s\right) + \frac{i\nu^2}{\kappa^J_n}M'_{-i\kappa^J_n,\mu}\left(\frac{i\nu^2}{\kappa^J_n}s\right) + \frac{\nu^2}{m}M_{-i\kappa^J_n,\mu}\left(\frac{i\nu^2}{\kappa^J_n}s\right),
\end{equation}
the eigenvalues $\kappa^J_n$ are determined through the relation
\begin{equation}
N_J(\kappa^J_n,\mu,1) D_J(\kappa^J_n,\mu,D) = N_J(\kappa^J_n, \mu, D) D_J(\kappa^J_n, \mu,1).
\end{equation}
Subsequently, the constant $\eta^J_n$ in equation~(\ref{jacket.modes}) is obtained:
\begin{equation}
\eta^J_n = \frac{N_J(\kappa^J_n,\mu,1)}{D_J(\kappa^J_n,\mu,1)} = \frac{N_J(\kappa^J_n,\mu,D)}{D_J(\kappa^J_n,\mu,D)}.
\end{equation}

\subsection{$p$-mode eigenvalues}
 The functional form of $p$ modes, given by equation~(\ref{modes.eigfunc}), has to satisfy
\begin{equation}
\frac{\partial\Phi_p}{\partial s} + \frac{\nu^2}{m}\Phi_p = 0,
\end{equation}
at $s = 1, D$. Following the formalism of appendix~\ref{jack.sec}, we define $N_p, D_p$ as:
\begin{equation}
N_p(\kappa^p_n,\mu,s) = \frac{1/2 + \mu}{s} M_{\kappa^p_n,-\mu}\left(\frac{\nu^2}{\kappa^p_n}s\right) - \frac{\nu^2}{\kappa^p_n}M'_{\kappa^p_n,-\mu}\left(\frac{\nu^2}{\kappa^p_n}s\right) -\frac{\nu^2}{m} M_{\kappa^p_n,-\mu}\left(\frac{\nu^2}{\kappa^p_n}s\right),
\end{equation}
and
\begin{equation}
D_p(\kappa^p_n,\mu,s) = -\frac{1/2 + \mu}{s} M_{\kappa^p_n,\mu}\left(\frac{\nu^2}{\kappa^p_n}s\right) + \frac{\nu^2}{\kappa^p_n}M'_{\kappa^p_n,\mu}\left(\frac{\nu^2}{\kappa^p_n}s\right) + \frac{\nu^2}{m}M_{\kappa^p_n,\mu}\left(\frac{i\nu^2}{\kappa^p_n}s\right),
\end{equation}
and determine the eigenvalue $\kappa^p_n$ and constant $\zeta^p_n$ in equation~(\ref{modes.eigfunc}) through the following relations, respectively:
\begin{equation}
N_p(\kappa^p_n,\mu,1) D_p(\kappa^p_n,\mu,D) = N_p(\kappa^p_n, \mu, D) D_p(\kappa^p_n, \mu,1),
\end{equation}
\begin{equation}
\zeta^p_n = \frac{N_p(\kappa^p_n,\mu,1)}{D_p(\kappa^p_n,\mu,1)} = \frac{N_p(\kappa^p_n,\mu,D)}{D_p(\kappa^p_n,\mu,D)}.
\end{equation}

\section{Determining the amplitude and phase of the scatter from observations}\label{howtogetit}
Through analyses of Michelson Doppler Imager (MDI) data \citep{scherrer} and the application of the forward models of \citet{gizon02}, \citet{duvall06} estimated the complex scattering amplitude ($d$ in their paper) due to network magnetic magnetic elements. Assume an incident plane wave of the form, i.e. an alternate form equation~(\ref{incident}),
\begin{equation}
\Phi^{\rm in} = e^{i k^p_0 x - i \omega t }\Phi_p(k^p_0;s),
\end{equation}
where $x = \xhat\cdot \br$, with the velocity of the wave given by,
\begin{equation}
\vel^{\rm in}_{\rm h} = \frac{\partial}{\partial t}\bnabla\Phi^{\rm in} = \omega \bk\; \Phi^{\rm in}.\label{vels.eqn}
\end{equation}
For dipole scattering only, the scattered $f$ mode takes on the form:
\begin{equation}
\Phi^{\rm sc} = -2 i \alpha^{p}_{10} \cos{\theta} \;  H^1_1(k_0 r)\Phi_p(k^p_0;s).\label{scat.eq}
\end{equation}
Using conventions from \citet{gizon02} and the definition of dipole scattering of \citet{duvall06}, we have,
\begin{equation}
{\mathcal L}_0 \delta v_z = - \delta{\mathcal L} \vel^{\rm in} = - \bnabla_h \cdot \left[ d \vel^{\rm in}_{\rm h} \right].\label{defns.many}
\end{equation}
Solving equation~(\ref{defns.many}) for a point scatterer,
\begin{equation}
\delta v_z(\br,\omega) = - 2\pi \int\id\bs\; G_z(\br-\bs,\omega) \bnabla_\bs\cdot \left[ d \delta(\bs) \; \vel^{\rm in}_{\rm h}(\bs,\omega)\right],
\end{equation}
which upon integrating by parts produces
\begin{equation}
\delta v_z(\br,\omega) = 2\pi d \int\id\bs\; \delta(\bs) \vel^{\rm in}_{\rm h}(\bs,\omega)\cdot\bnabla_\bs G_z(\br-\bs,\omega),
\end{equation}
or,
\begin{equation}
\delta v_z(\br,\omega) = - 2\pi d \; \vel^{\rm in}_{\rm h}(0,\omega)\cdot\bnabla_{\br} G_z(\br,\omega).\label{eqn.greens}
\end{equation}
In the far field, the Green's function reduces to
\begin{equation}
G_z(\br) \approx \frac{i}{4\pi g} H_0^1(k_0 r).
\end{equation}
The derivative in equation~(\ref{eqn.greens}) becomes,
\begin{equation}
\bnabla_{\br} G_z(\br) \approx -\frac{i k}{4\pi g} H_1^1(k_0 r) \hat{\br},
\end{equation}
and equation~(\ref{eqn.greens}) along with the expression for the wave velocity from equation~(\ref{vels.eqn}) simplifies to
\begin{equation}
\delta v_z(\br,\omega) = \Psi_p(k^p_0;s)  \frac{i \omega k^2 d}{2 g} H_1^1(k_0 r) \cos\theta,
\end{equation}
and therefore,
\begin{equation}
\Phi^{\rm sc} = \delta \Phi = \delta v_z / (-i k \omega) =  - \frac{k d}{2 g} \cos\theta  H_1^1(k_0 r) \Phi_p(k^p_0;s).
\end{equation} 
From equation~(\ref{scat.eq}), we obtain the following scattering coefficient: 
\begin{equation}
\alpha^p_{10} = - \frac{i k}{4 g} d
\end{equation}
For $d$ with a phase of about 130 deg \citep{duvall06}, $\alpha^p_{10}$ has a phase of about 40 deg.

\end{document}